\documentclass[fleqn,10pt,sort&compress]{wlscirep}
\usepackage[numbers]{natbib}
\usepackage{graphicx}
\usepackage{subcaption}

\usepackage[T1]{fontenc}
\usepackage[ttdefault=true]{AnonymousPro}
\usepackage{url} 
\usepackage{hyperref} 

\usepackage[T1]{fontenc}
\usepackage[utf8]{inputenc}
\usepackage[scaled]{helvet}
\usepackage[mathletters]{ucs}

\usepackage{textcomp}  

\usepackage{gensymb}   

\newcommand{\msun}{{\rm M}_{\odot}}
\newcommand{\beq}{\begin{equation}}
\newcommand{\eeq}{\end{equation}}
\newcommand{\bea}{\begin{eqnarray}}
\newcommand{\eea}{\end{eqnarray}}

\newenvironment{cititemize2}
{\begin{list}{$\bullet$}
        {\setlength{\topsep}{0pt}
         \setlength{\itemsep}{0pt}
         \setlength{\parsep}{0.25\parsep}
         \settowidth{\labelwidth}{$\bullet$}
         \setlength{\leftmargin}{1em}
}
}
{\end{list}}

\begin{document}

\title{Accelerated, Scalable and Reproducible AI-driven Gravitational Wave Detection}

\author[1,2,*]{E. A. Huerta}
\author[3]{Asad Khan}
\author[3]{Xiaobo Huang}
\author[3]{Minyang Tian}
\author[2]{Maksim Levental}
\author[1]{Ryan Chard}
\author[3]{Wei Wei}
\author[3]{Maeve Heflin}
\author[3]{Daniel S. Katz}
\author[3]{Volodymyr Kindratenko}
\author[3]{Dawei Mu}
\author[1,2]{Ben Blaiszik}
\author[1,2]{Ian Foster}

\affil[*]{e-mail: elihu@anl.edu}

\affil[1]{Data Science and Learning Division, Argonne  National  Laboratory,  Lemont,  Illinois  60439,  USA}
\affil[2]{University of Chicago, Chicago,  Illinois  60637,  USA}
\affil[3]{University of Illinois at Urbana-Champaign, Urbana, Illinois 61801, USA}

\begin{abstract}
\noindent The development of reusable artificial intelligence (AI) models for wider use and 
rigorous validation by the community promises to unlock new opportunities in 
multi-messenger astrophysics. Here we develop a workflow that connects the 
\texttt{Data and Learning Hub for Science}, a repository for publishing AI models,  with 
the Hardware Accelerated Learning (\texttt{HAL}) cluster, using \texttt{funcX} as a 
universal distributed computing service. Using this workflow, an ensemble of four openly 
available AI models can be run on \texttt{HAL} to process an entire month's worth (August 2017) 
of advanced Laser Interferometer Gravitational-Wave Observatory data in just seven minutes, 
identifying all four all four binary black hole mergers previously identified in this dataset and 
reporting no misclassifications. This approach combines advances in AI, distributed computing, 
and scientific data infrastructure to open new pathways to conduct reproducible, accelerated, 
data-driven discovery.

\end{abstract}

\flushbottom
\maketitle

\section*{Introduction}

Gravitational waves were added to the growing set of detectable cosmic 
messengers in the fall of 2015 when the advanced Laser Interferometer 
Gravitational-Wave Observatory (LIGO) detectors 
reported the observation of gravitational waves consistent with 
the collision of two massive, stellar-mass black holes~\cite{DI:2016}. 
Over the last five years, the advanced LIGO and advanced Virgo 
detectors have completed three observing runs, reporting over 
50 gravitational wave sources~\cite{PhysRevX.9.031040,Abbott:2020niy}. 
As advanced LIGO and advanced Virgo continue to 
enhance their detection capabilities, and other detectors join the international 
array of gravitational wave detectors, it is 
expected that gravitational wave sources will be observed at a rate of several per day~\cite{Abbott:2020gyp}. 

An ever-increasing catalog of gravitational waves will 
enable systematic studies to advance our 
understanding of stellar evolution, cosmology, alternative 
theories and gravity, the nature of supranuclear matter in neutron stars, and the formation and 
evolution of black holes and neutron stars, among 
other phenomena~\cite{Soares-Santos:2019,Abbott,Hubble_from_GW,Berti:2018GReGr,LIGOScientific:2019fpa,Radice:2016MNRAS,Metzger:2019zeh}. 
Although these science goals are feasible in principle given 
the proven detection capabilities of astronomical observatories, 
it is  equally true that established algorithms for 
the observation of multi-messenger sources,
such as template matching and nearest neighbors,
are compute-intensive 
and poorly scalable~\cite{HuertaBWOSG,HuertaES,2017Weitzel}. Furthermore, available computational resources will remain oversubscribed, and planned enhancements will be rapidly outstripped with the advent of next-generation detectors within the next couple of years~\cite{HPCUSE}. Thus, an urgent re-thinking is critical if we are to realize 
the multi-messenger astrophysics program in the big-data era~\cite{2019NatRP...1..600H}. 

To contend with these challenges, a number of researchers have been exploring 
the application of deep learning and of computing accelerated with graphics 
processing units (GPUs). Co-authors of this article pioneered the use of 
deep learning and high performance computing to accelerate the detection of 
gravitational waves~\cite{geodf:2017a,GEORGE201864}. The first generation of these 
algorithms targeted a shallow signal manifold (the masses of the binary components) 
and only required tens of thousands of modeled waveforms for training, 
but these models served the purpose of demonstrating that an alternative 
method for gravitational wave detection is as 
sensitive as template matching and significantly faster, at a fraction of the computational cost.

Research and development in deep learning is moving at an incredible pace~\cite{2018GN,Lin:2020aps,Wang:2019zaj,2017CQGra..34f4003Z,remove_blip,Nakano:2018vay,Fan:2018vgw,Deighan:2020gtp,Miller:2019jtp,Krastev:2019koe,2020PhRvD.102f3015S,Khan:2020foe,Dreissigacker:2019edy,2020PhRvD.101f4009B,2020arXiv200914611S,Khan:2020fso,PhysRevLett.122.211101,wei_warning,wei_ecc_princ} (see also ref.~\cite{cuoco_review} for a review of 
machine-learning applications in gravitational wave astrophysics). Specific milestones in 
the development of artificial intelligence (AI) tools for gravitational wave astrophysics include 
the construction of neural networks that describe the four-dimensional (4D) signal manifold 
of established gravitational wave detection pipelines, that is, the masses of the binary 
components and the \(z\)-component of the three-dimensional spin vector 
in \((m_1, m_2, s_1^z, s_2^z)\). This requires the combination of distributed training 
algorithms and extreme-scale computing to train these AI models with 
millions of modeled waveforms in a reasonable amount of 
time~\cite{Khan:2020foe}. Another milestone concerns the creation of AI models that 
enable gravitational wave searches over hour-long datasets, keeping the number 
of misclassifications at a minimum~\cite{Wei_Khan_Huerta}. 

In this article, we introduce an AI ensemble, 
designed to cover the 4D signal manifold \((m_1, m_2, s_1^z, s_2^z)\), to search 
for and find binary black hole mergers over the entire month of August 2017 in 
advanced LIGO data~\cite{Vallisneri:2014vxa}. Our findings indicate that this approach 
clearly identifies all black hole mergers contained in that data batch 
with no misclassifications. To conduct this analysis we used the Hardware-Accelerated 
Learning (\texttt{HAL}) cluster deployed and operated by the 
Innovative Systems Lab at the National Center for Supercomputing Applications. 
This cluster consists of 16 \texttt{IBM SC922 POWER9} nodes, 
with four \texttt{NVIDIA} V100 GPUs per node~\cite{HAL2020}. 
The nodes are interconnected with \texttt{EDR} \texttt{InfiniBand} network, 
and the storage system is made of two \texttt{DataDirect Networks} all-flash arrays 
with \texttt{SpectrumScale} file system, providing 250 TB of usable space. 
Job scheduling and resources allocation are managed by the \texttt{SLURM} (Simple 
Linux Utility for Resource Management) system. As we show below, 
we can process the entire month of August 2017 with our deep learning ensemble in just 
7 min using the entire \texttt{HAL} cluster. In addition to taking this 
noteworthy step forward in the use of AI for accelerated gravitational wave searches, 
we also demonstrate that we can 
share these models with the broader community by leveraging the 
Data and Learning Hub for Science (\texttt{DLHub})~\cite{li147dlhub,dlhub8821027}. 
We believe that this 
approach will accelerate the adoption and further development of 
deep learning for gravitational wave astrophysics.

Given that \texttt{DLHub} has the ability to both archivally store and actively run 
trained models, it provides a means to address reproducibility, reuse, and credit. 
With sufficient computational resources (\texttt{HAL}, in this case) and a connection 
made via \texttt{funcX}, a function-as-a-service platform, having a model on 
\texttt{DLHub} allows the inference or analysis described in a published paper to be 
reproduced given that the original data is also available (as described in Results). 
If applied to new data instead, the model can then be reused, with \texttt{DLHub}'s 
registration providing a means for the developers of the model to receive credit 
by the users citing the archived model. This also allows users to easily experiment 
with trained models, even from one discipline to another.

This paper brings together several key elements to accelerate deep learning research. 
We showcase how to combine cyberinfrastructure funded by the National Science 
Foundation (NSF) and the Department of Energy (DOE) to release 
state-of-the-art, production scale, neural network models for gravitational wave 
detection. The framework  \(\texttt{DLHub}\rightarrow\texttt{funcX}\rightarrow\texttt{HAL}\) 
provides the means to enable open source, accelerated deep learning gravitational 
wave data analysis. This approach will empower the broader community to readily 
process open source LIGO data with minimal computational resources. Going forward, 
this approach may be readily adapted to demonstrate interoperability, replacing \texttt{HAL} 
with any other compute resource. At a glance, the developments for AI-driven 
gravitational wave detection introduced in this article encompass the following 
characteristics:

\begin{cititemize2}
\item \textbf{Open Source} The containerized AI models introduced in this study are 
shared with the broader community through \texttt{DLHub}. This approach will 
streamline and accelerate the adoption and development of AI for gravitational 
wave astrophysics
\item \textbf{Reproducible} We conducted two independent analyses to test the 
predictions of our AI ensemble, and confirm that the output of these studies 
is consistent and reproducible 
\item \textbf{Accelerated} We used 64 \texttt{NVIDIA} GPUs to process 
advanced LIGO data from the entire month of August 2017 in just 7 min, which is 
orders of magnitude faster and computationally more efficient than other 
methods that have harnessed 
advanced cyberinfrastructure platforms for gravitational wave 
detection~\cite{HuertaBWOSG,2017Weitzel}. 
\item \textbf{Sensitivity and Accuracy} This data-driven approach is 
capable of processing advanced LIGO data in bulk for a 4D signal manifold, 
reporting perfect true positive rate on real gravitational wave events and 
zero misclassifications over one month's worth of searched data
\item \textbf{Scalable} We demonstrate that AI-driven gravitational wave detection 
scales strongly as we increase the number of GPUs used for inference. 
The software needed to scale this analysis, and to post-process the output of the 
AI ensemble is all provided at the \texttt{DLHub}
\end{cititemize2}

\noindent The outstanding aspect of this work is the consolidation of these five disparate 
elements into a unified framework for end-to-end AI-driven gravitational wave detection. 
This type of big-data, open science research is part of a global project that aims to 
harness AI and advanced cyberinfrastructure to 
enable innovation in data-intensive research through new modes of data-driven discovery. 
Examples are the NSF Harnessing the Data Revolution and DOE FAIR (Findable, Accessible, 
Interoperable, and Reusable) Projects in the United States; and the European Science 
Cluster of Astronomy \& Particle Physics ESFRI Research Infrastructures (ESCAPE) 
Project; and the European Open Science Cloud (EOSC)~\cite{2020arXiv201211534A}.

\section*{Results}
\label{sec:result}

We present results for two types of analyses. The first set of results was 
obtained by running our AI ensemble directly in \texttt{HAL} on advanced LIGO 
noise. For the second set, we conducted a similar analysis but now using the AI 
ensemble hosted at the \texttt{DLHub} and connecting it to \texttt{HAL} through 
\texttt{funcX}. These analyses were independently carried out by two different teams.

\begin{figure*}[!htb]
\centerline{
\includegraphics[width=\textwidth]{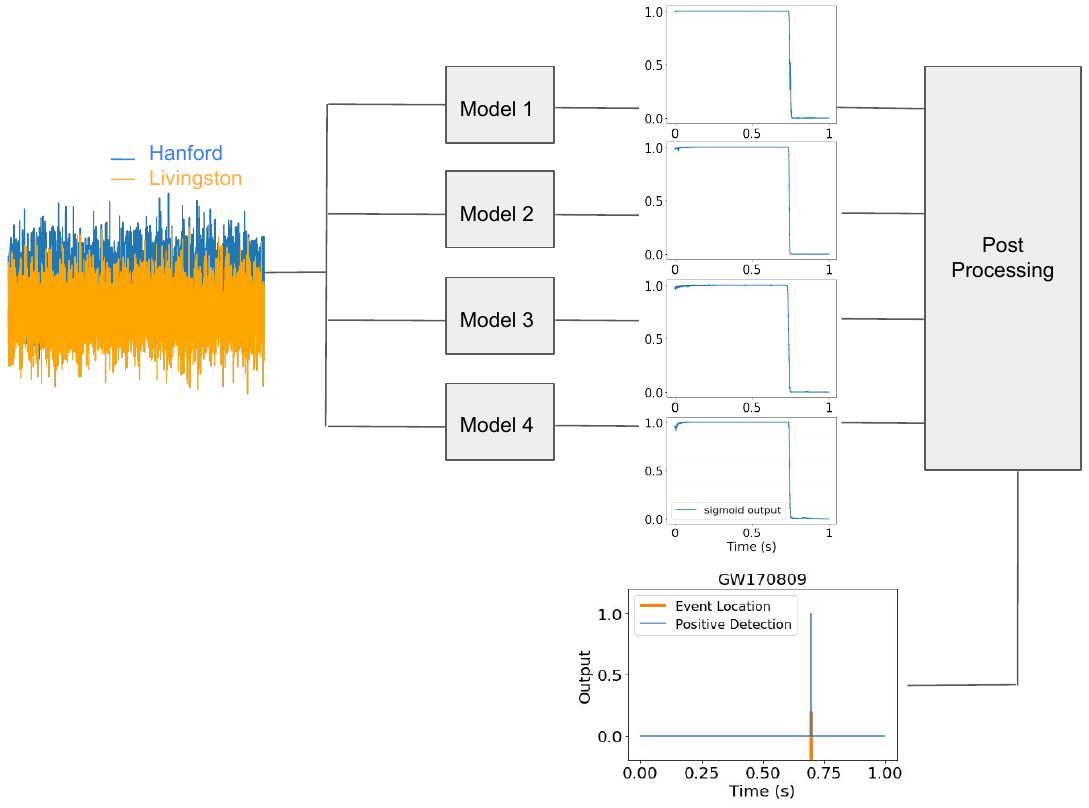}
} 
\caption{\textbf{Gravitational wave detection workflow with AI ensemble.} Hanford and 
Livingston gravitational wave data, depicted as blue and orange time-series data on 
the left, are fed into an AI ensemble of four neural network models. The 
response of the neural networks to advanced 
LIGO data is shown to the right of the boxes representing the models. 
At the post-processing stage, the outputs of the four neural networks are combined. 
If the outputs of all the models are consistent with the existence of a gravitational wave signal, 
then the post-processing algorithm indicates a positive detection. The bottom panel 
showcases a positive detection for the binary black hole merger  GW170809.}
\label{fig:workflow}
\end{figure*}

\paragraph{Accelerated AI gravitational wave detection in \texttt{HAL}.}
We designed an AI ensemble, described in detail in Methods  
and schematically shown 
in Fig.~\ref{fig:workflow}, to process advanced 
LIGO open-source data, from both Livingston and Hanford, that cover 
the entire month of August 2017. 
We chose this month because it contains several gravitational wave sources, 
and thus it provides a test bed to quantify the sensitivity of our AI models in 
identifying real events and to estimate the number of false positives 
over extended periods of time.

\begin{figure}[!htb]
\centerline{
\includegraphics[width=\textwidth]{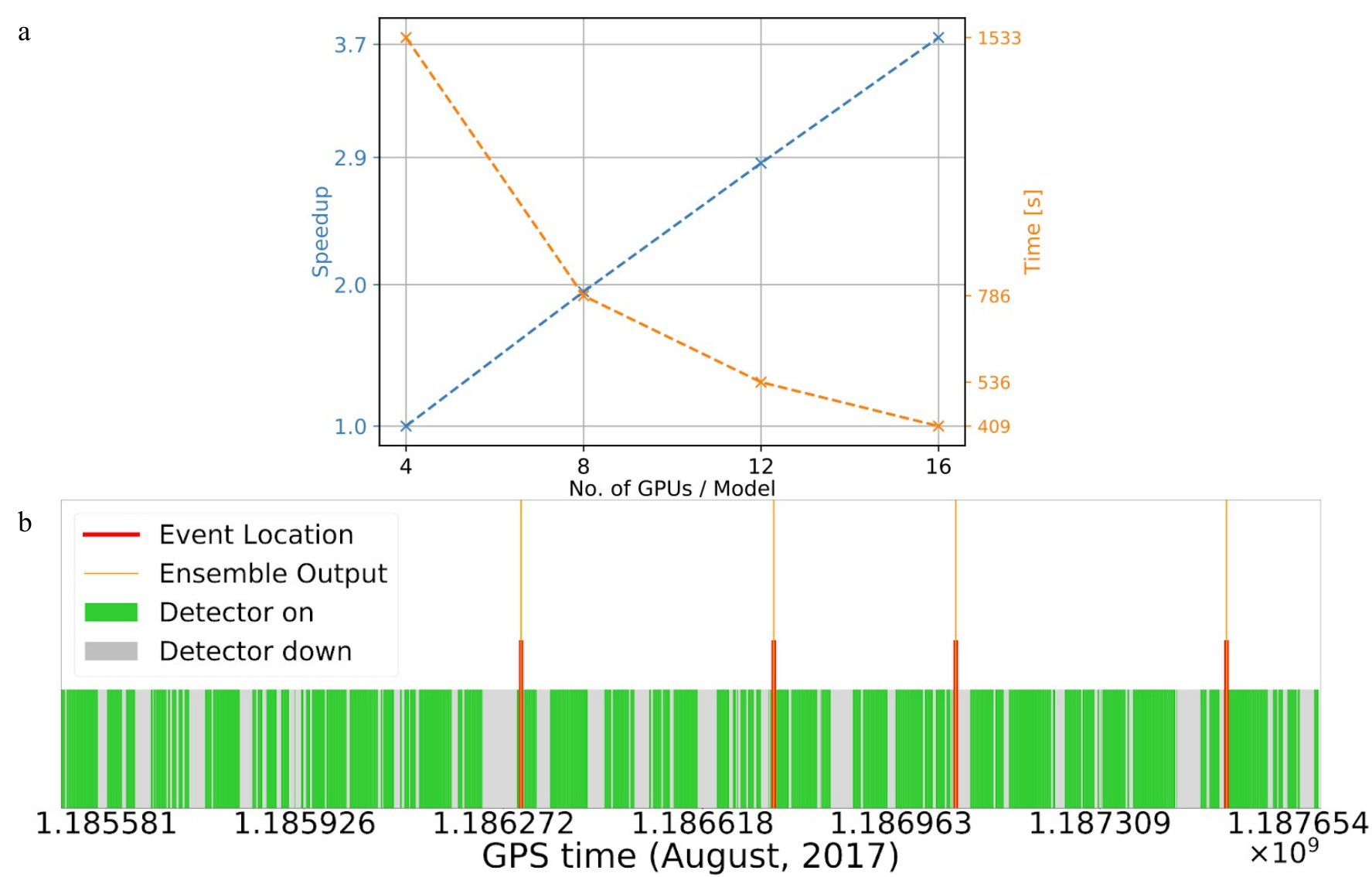}
} 
\caption{\textbf{Speed, sensitivity and scalability of AI ensemble.}  \textbf{a,} The blue line 
indicates that we obtain near-perfect scaling as we distribute our AI ensemble over 
the entire HAL cluster. The orange line shows that our AI ensemble may process Hanford 
and Livingston datasets that span August 2017 in about 25 minutes when each neural network 
in the ensemble is assigned four \texttt{NVIDIA} V100 GPUs. Assigning 16 V100 GPUs to 
each model in the ensemble reduces the gravitational wave search to just 7 min. \textbf{b,} The 
green segments indicate the times when both Hanford and Livingston detectors were 
collecting data. Grey lines show times when one or both detectors were down. The orange 
lines show the output of our AI ensemble, which coincides with the existence of real 
gravitational waves (indicated by red lines) of binary black hole mergers in August 2017.}
\label{fig:ensemble_long}
\end{figure}

\begin{figure}[!htb]
\centerline{
\includegraphics[width=\textwidth]{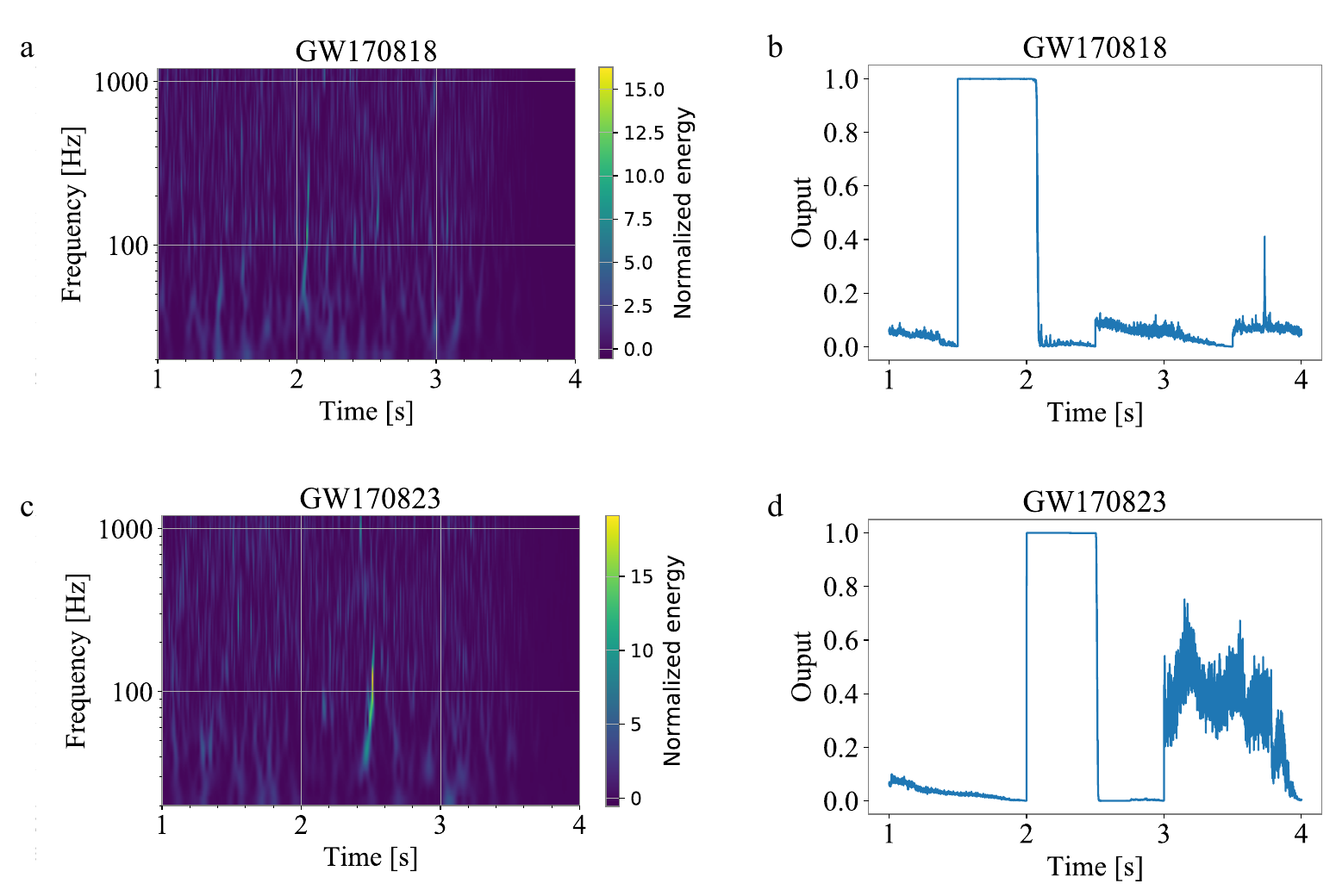}
}
\caption{\textbf{Spectrograms and neural network response to gravitational waves.} \textbf{a,c,} 
L-channel spectrograms of gravitational wave sources identified by our AI ensemble. \textbf{b,d,} 
The signals in \textbf{a} and \textbf{c} produce a corresponding sharp, distinctive response in our neural network models.}
\label{fig:follow_up}
\end{figure}

The results we present below were obtained by directly running 
our AI ensemble in \texttt{HAL}. Fig.~\ref{fig:ensemble_long} 
summarizes the speed and sensitivity of our approach. When we distributed the 
inference over all 64 \texttt{NVIDIA} V100 GPUs in \texttt{HAL}, we can complete 
the search over the entire month of August 2017, including 
post-processing of the noise triggers identified by each model 
in the ensemble, within just 7 min (Fig. 2a). Our AI ensemble identifies all four binary 
black hole mergers contained in this dataset (Fig. 2b). We follow up two 
of these gravitational wave sources in Fig.~\ref{fig:follow_up}, which 
presents spectrograms (Fig. 3a, c) and the response of one of the AI models 
in the ensemble to these real events (Fig. 3b, d). 

\begin{figure}[!htb]
    \centering
    \includegraphics{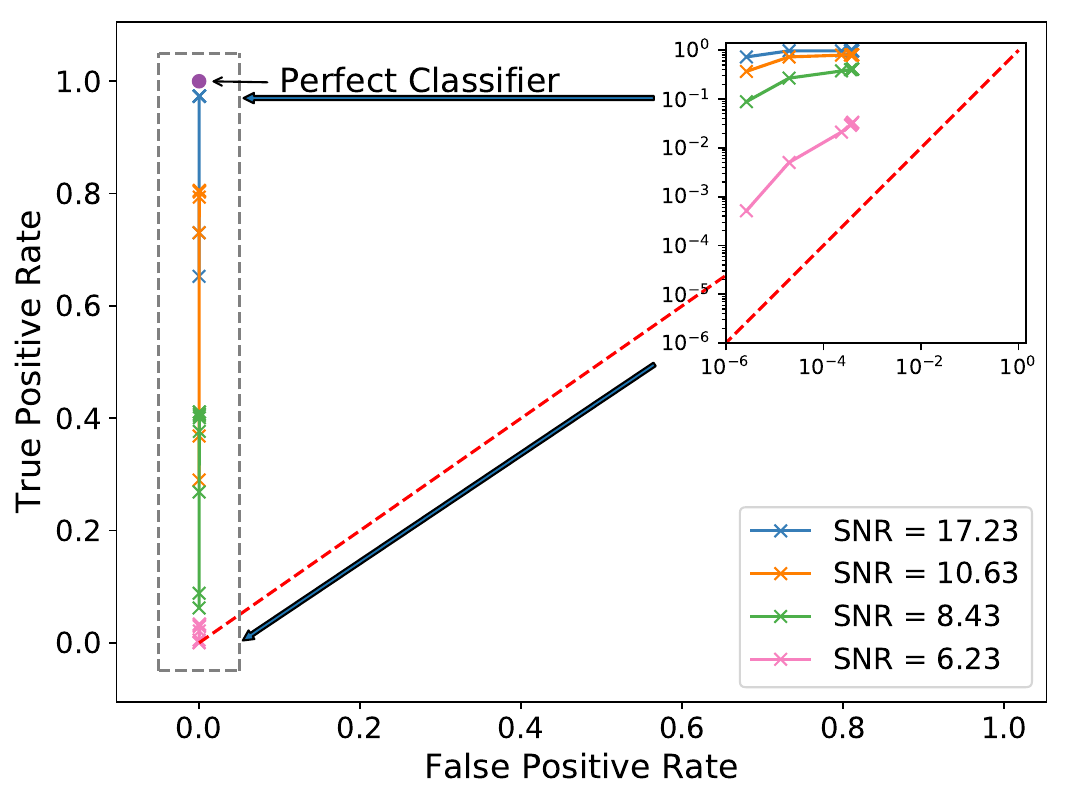}
    \caption{\textbf{Receiver operating characteristic curve of AI ensemble.}  The lines show 
    the ROC curves for a test set that contain 237,663 modeled binary black hole 
    waveforms injected in advanced LIGO noise throughout August 2017 and that covers a 
    broad SNR range. The true positive rate is shown against the false positive rate as 
    estimated from the output of our ensemble of four AI models. For reference, we indicate 
    the performance of a ``perfect classifier"  in the top-left corner, i.e., 100\% sensitivity with 
    no false positives. The red dashed line describes the performance of an untrained model 
    that produces random guesses. The grey dotted lines indicate the region of the inset.}
    \label{fig:roc_curve}
\end{figure}

We quantified the performance of our AI ensemble for gravitational wave detection 
by computing the receiver operating characteristic (ROC) curve using a test set of 
237,663 modeled waveforms, injected in advanced LIGO noise and that cover a 
broad signal-to-noise ratio (SNR) range. 
As described in detail in Methods, we post-process the output of our AI ensemble 
with the \texttt{find\_peaks} algorithm so that the width of the peak is within  
$[0.5,2]$ seconds and the height is between 0 and 1. In Fig.~\ref{fig:roc_curve} 
we vary the height threshold between 0 and 1 while maintaining a minimum width of 
the peak to 0.5s. Our AI ensemble attains optimal performance
 in true positive rate as we increase the threshold from 0 to 0.9998, while 
 the false positive rate increases from  $10^{-6}$ to $10^{-3}$. 
 The AI approach achieves a high level of sensitivity and computational performance 
over long stretches of real advanced LIGO data. This approach is also 
capable of an accelerated, reproducible, AI gravitational wave search at scale 
that covers the 4D signal manifold that describes quasi-circular, spinning, 
non-precessing binary black hole mergers.

\begin{figure}[!htb]
    \centering
    \includegraphics[width=\textwidth]{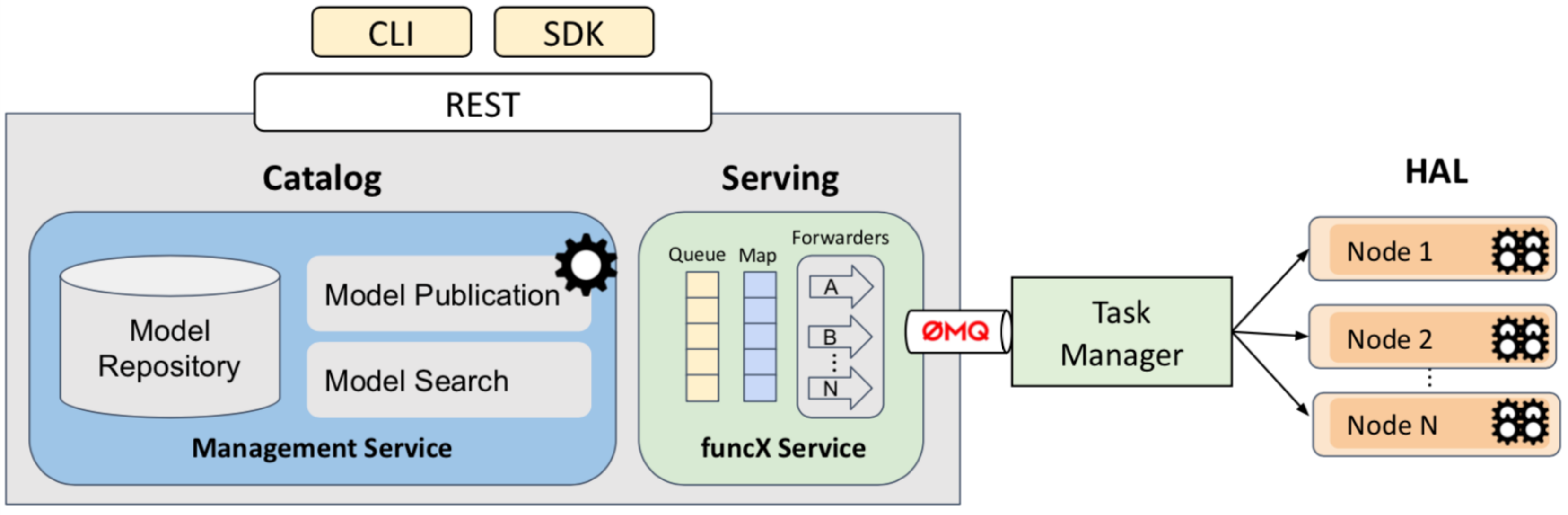}
    \caption{\textbf{\texttt{DLHub} architecture.} Schematic representation of the 
    cyberinfrastructure resources used to conduct accelerated and reproducible gravitational 
    wave detection on open-source advanced LIGO data. This architecture provides a 
    command line interface (CLI), a Python software development kit (SDK) and a 
    representational state transfer (REST) application programming interface to publish, 
    manage and invoke AI models. The management service coordinates the execution of 
    tasks on remote resources using a ZeroMQ ($\texttt 0$MQ) queue, which sends tasks 
    to registered task managers for execution. This messaging model ensures that 
    tasks are received and executed. \texttt{DLHub} supports both synchronous and 
    asynchronous task execution.}
    \label{fig:dlhub}
\end{figure}

\paragraph{Connection of \texttt{DLHub} to \texttt{HAL} through \texttt{funcX}.}
We exercise the AI ensemble through \texttt{DLHub}~\cite{dlhub8821027} 
so that the models are widely available and our results reproducible.
\texttt{DLHub} is a system that provides model repository and model serving facilities, 
in particular for machine-learning models with science applications. 
\texttt{DLHub} itself leverages \texttt{funcX}~\cite{chard20funcx}, a function-as-a-service 
platform that enables high-performance remote function execution in a flexible, scalable, 
and distributed manner. A \texttt{funcX} endpoint is deployed to \texttt{HAL} and registered 
with the \texttt{funcX} service; an endpoint consists of a \texttt{funcX} \texttt{agent}, a 
\texttt{manager} and a \texttt{worker}, and abstracts the underlying computing resource. 
The endpoint dynamically provisions \texttt{HAL} nodes and deploys 
\texttt{workers} to perform inference requests. This computational infrastructure 
is schematically shown in Fig.~\ref{fig:dlhub}. 

\begin{figure}[!htb]
    \centering
    \includegraphics[width=0.7\textwidth]{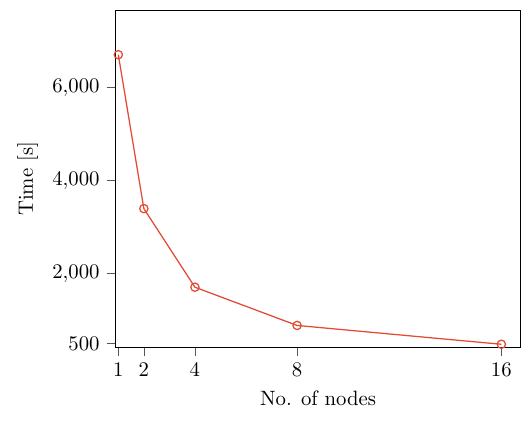}
    \caption{\textbf{Throughput of the \texttt{DLHub+HAL} architecture.}  An AI ensemble of 
    four neural networks, hosted at the \texttt{DLHub}, processes advanced LIGO data from 
    an entire month (August 2017) in 7 min using the 
    entire \texttt{HAL} cluster that has 64 \texttt{NVIDIA} V100 GPUs evenly distributed
     over 16 nodes.}
    \label{fig:dlhub_exec}
\end{figure}

\texttt{DLHub} packages the AI ensemble and the inference function as a 
\textit{servable}, that is, it builds a container that includes the models and 
their dependencies and registers the inference 
function with the \texttt{funcX} registry. 
An inference run corresponds to a \texttt{funcX} function invocation, 
which triggers the dispatch of 
the task to the \texttt{HAL} \texttt{funcX} endpoint.
On every such run (that is, per collection of data, not per sample) the 
\texttt{funcX} agent deployed to 
\texttt{HAL} allocates the task to multiple nodes and managers (one manager per node), 
each of which then forwards the task to a single worker, which finally distributes the 
task across the GPUs in a multiple instruction, single data fashion (that is, one model 
per GPU, with all models operating on the same data).
The models then perform their individual forward passes and the workers 
aggregate their results.
Note that distinct managers, and therefore workers on distinct nodes,  
operate on non-overlapping chunks of data. 
Fig.~\ref{fig:dlhub_exec} shows that scaling the inference analysis through 
the \texttt{DLHub} architecture is equivalent to directly running the analysis on 
the \texttt{HAL} cluster through distributed inference. In both instances, we succeeded 
at using the entire \texttt{HAL} cluster optimally.

Using this computational infrastructure, we reproduced the 
results presented in the previous section. Not only that, 
we also found that the computational performance of 
the \texttt{DLHub} architecture provides the same throughput 
as running the AI ensemble directly in \texttt{HAL}.

Currently the system is set up only for post-processing of data, 
which provides the required 
framework for accelerated, off-line analyses. We are working to extend 
\texttt{DLHub} and \texttt{funcX} 
to support streaming data by using Globus~\cite{globus7036262}. This approach is 
laying the foundations for future scenarios in which advanced LIGO 
and other astronomical observatories broadcast real-time data 
to the broader community. Such an approach would enable researchers to carry 
out analyses with open-source data that are beyond the scope of 
scientific collaborations but essential to push the 
frontiers of multi-messenger astrophysics.

\section*{Discussion}

Innovative AI applications in gravitational wave astrophysics have evolved 
from disruptive AI prototypes~\cite{geodf:2017a,GEORGE201864} into sophisticated, 
physics-inspired AI 
models that describe the signal manifold covered by traditional gravitational wave detection 
pipelines for binary black hole mergers~\cite{Wei_Khan_Huerta}. These models have 
the same sensitivity as 
template matching algorithms and run orders of magnitude faster and at a fraction 
of the computational cost. 

AI is now being used to accelerate the 
detection of binary black holes and binary neutron 
stars~\cite{Wei_Khan_Huerta,wei_warning,Krastev:2019koe,2020PhRvD.102f3015S}, 
and to forecast the 
merger binary neutron stars and neutron star-black 
hole systems~\cite{wei_warning,wei_ecc_princ}. The current pace of progress makes 
it clear that the broader community 
will continue to advance the development of AI to realize the 
science goals of multi-messenger astrophysics.

Mirroring the successful approach of corporations leading AI 
innovation, we are releasing our AI models to 
enable the broader community to use and perfect them. This approach is also 
helpful to address 
healthy and constructive skepticism from researchers who do not feel at ease using 
AI. This article also demonstrates how complementary communities can 
work together to harness DOE- and NSF-funded cyberinfrastructure 
to enable accelerated, reproducible, AI-driven,  
compute-intensive analyses in record time. This approach will 
facilitate a plethora of new 
studies since \texttt{DLHub} and 
\texttt{funcX} are discipline-agnostic and hardware-agnostic.

\section*{Methods}
\label{sec:meth}

\noindent \textbf{Data.} Datasets used to train, validate and test our AI ensemble are open source and readily accessible as described below.
\\

\noindent \textbf{Modeled Waveforms} We used the open source \texttt{PyCBC} library~\cite{pycbc_library} 
to produce modeled waveforms 
with the approximant \texttt{SEOBNRv3}~\cite{seobnrv3}. The training dataset 
consists of 1,136,415 modeled waveforms, sampled at 4096 Hz, that cover a 
parameter space of binary black hole mergers with 
total masses \(M\in[5\msun,\,100\msun]\), mass ratios \(q \leq 5\), and individual 
spins \(s^z_{\{1,2\}}\in[-0.8,\,0.8]\). 
\\

\noindent \textbf{Advanced LIGO gravitational wave data.} The modeled waveforms 
are whitened and linearly mixed with advanced LIGO noise obtained from the 
\texttt{Gravitational Wave Open Science Center}~\cite{Vallisneri:2014vxa}. 
Specifically, we use the three noise data segments, each 4096 s long, 
starting at GPS times $1186725888$, $1187151872$, 
and $1187569664$. 
None of these segments include known gravitational wave detections. 
These data are used to compute noise power spectral density estimates with open-source 
code available at the 
\texttt{Gravitational Wave Open Science Center}. These power spectral densities 
are then used to whiten both the strain data and the 
modeled waveforms. Thereafter, the whitened strain data and the whitened modeled 
waveforms are linearly 
combined, and a broad range of SNRs are covered to encode scale invariance in the 
neural network. We then normalize the standard deviation of training data 
that contain both signals and 
noise to one. The ground-truth labels for training are encoded such that each time 
step after the merger is classified as noise, and all preceding time-steps in the 1 second 
window are classified as waveform strain. Hence, the transition in classification 
from waveform to noise identifies the location of the merger. 
\\

\noindent \textbf{Code.} Our AI ensemble and the post-processing 
software used to search for and find gravitational waves are available 
at the DLHub~\cite{dlhubmodelgravs}. 
Each model in our AI ensemble consists of two independent modified 
\texttt{WaveNet}s~\cite{2016wavenet} processing Livingston and Hanford strain data 
sampled at 4,096Hz. The two outputs are then concatenated and jointly fed into a final set 
of two convolutional layers which output a classification (noise or waveform) probability 
for each time step.  At test time on advanced LIGO strain data, we employ a 
post-processing function to precisely locate such transitions in the model's output. 
Specifically, we use a 1 second window on the strain data, with a step size of 0.5 seconds, 
and use off-the-shelf peak detection algorithm \texttt{find\_peaks}, provided by \texttt{SciPy}, 
on the models probability output. In the \texttt{find\_peaks} algorithm, we specify the 
thresholds so that only peaks with a width in the 0.5 s--2 s range are selected. As 
the time step for the sliding window is 0.5 s, we merge any repeated detections, 
that is, peaks within 0.5 s of each other are counted as repeated detection. 
Once we have the predicted locations of the peaks or mergers from each of the four 
(randomly initialized and trained) models in the ensemble, we combine them in one 
final post-processing step so that all peaks that are within 
\(1/128\) s of each other are flagged as detection of true gravitational wave events, 
while the rest are discarded as random false alarms.
\\

\noindent \textbf{Statistics.} Using the aforementioned methodology, we have quantified 
the performance of our AI ensemble for classification (gravitational wave detection) by 
computing the ROC curve. For this calculation we used a test set that consists 
of 237,663 modeled waveforms that cover a broad SNR range. Note that we are 
able to reduce misclassifications by combining two methodologies. First, the use of four AI 
models in tandem enables us to discard noise anomalies that are flagged by only 
some of the models. For instance, we found in ref.~\cite{Wei_Khan_Huerta} that 
using two AI models still led to the misclassification of two loud noise anomalies 
as true gravitational wave signals. However, we have found that using four AI models 
removes these misclassifications, as some of the models in the ensemble did not 
flag these glitches as potential gravitational wave events. Second, we can calibrate 
the performance of the AI ensemble during training using long data segments. 
As mentioned above, in the post-processing stage we set the threshold of the 
\texttt{find\_peaks} algorithm so that the width of the peak is within the 0.5 s--2 s range 
and the height is between 0 and 1. To compute the ROC curve, we vary the height 
threshold between 0 and 1 while maintaining a minimum peak width of 0.5 s. With this 
approach, our AI ensemble attains optimal performance in true positive rate as 
we increase the threshold from 0 to 0.9998 while the false positive rate increases 
from  $10^{-6}$ to $10^{-3}$.

We notice that although our AI ensemble reduces significantly the number of 
misclassifications when processing advanced LIGO data in bulk, our methodology 
has room  for improvement. For instance, our AI ensemble is close but not identical to an optimal classifier, which we have marked in the top left corner of Fig.~\ref{fig:roc_curve}. 
Our vision to continue to improve the performance of AI models for gravitational 
wave detection includes the development of physics-inspired architectures 
and optimization schemes to enhance the sensitivity of AI ensembles; the incorporation 
of rapid regression algorithms that provide internal consistency checks on the nature of 
noise triggers, for example, independent estimation of the total mass of a potential 
binary system and the associated frequency at merger; and the inclusion of include 
open-source \texttt{GravitySpy}~\cite{2017CQGra..34f4003Z} glitches during the training 
stage to boost the ability of AI models to tell apart real signals from noise anomalies and 
more confidently identify real events. We sincerely hope that the methodology introduced 
in this article is used, improved, and extended by a broad set of users. 
Such an approach will lead to the development of increasingly better and more 
robust AI tools for data-driven discovery.

\section*{Data Availability}
Advanced LIGO data used in this manuscript is open-source and readily 
available at the \texttt{Gravitational Wave Open Science Center}~\cite{Vallisneri:2014vxa}. 
Modeled waveforms used to train, validate and test our AI models were produced using 
the open-source \texttt{PyCBC} library~\cite{pycbc_library}. The waveform family used 
was \texttt{SEOBNRv3}~\cite{seobnrv3}. The datasets generated and/or analysed during the 
current study are available from the corresponding author upon reasonable request.

\section*{Code Availability} 
All the required software to reproduce our results, encompassing AI models 
and post-processing scripts, are open source and readily available at the \texttt{DLHub} 
and may be found at~\cite{dlhubmodelgravs}. Software to produce waveforms at scale in 
high-performance computing platforms with \texttt{PyCBC} may be provided upon request.

\bibliography{mybibfile}

\section*{Acknowledgements}
\noindent We gratefully acknowledge National Science Foundation (NSF) awards 
OAC-1931561 and OAC-1934757 (E.A.H.),  OAC-1931306 (B.B.) and OAC-2004894 (I.F.).  
E.A.H. gratefully acknowledges the Innovative and Novel 
Computational Impact on Theory and Experiment project `Multi-Messenger Astrophysics 
at Extreme Scale in Summit'. This research used resources of the Oak Ridge 
Leadership Computing Facility, which is a DOE Office of Science User Facility supported 
under contract no. DE-AC05-00OR22725. 
This work utilized resources supported by the NSF's Major Research Instrumentation 
program, the HAL cluster (grant no. OAC-1725729), 
as well as the University of Illinois at Urbana-Champaign. \texttt{DLHub} is based upon work 
initially supported by Laboratory Directed Research and Development funding 
from Argonne National Laboratory, provided by the Director, Office of Science, of the DOE 
under contract no. DE-AC02-06CH11357.  We thank \texttt{NVIDIA} for their continued support.

\section*{Author Contributions}
E.A.H. led this work and coordinated the writing of this manuscript. A.K. developed and 
trained the AI ensemble, as well as the software to scale this analysis over the entire HAL 
cluster, and to post-process the output of the AI ensemble to estimate sensitivity and 
error analysis. X.H., M.T., M.H. and W.W. prepared the datasets and software used 
for training, testing and inference and conducted an independent inference study to 
ascertain the reproducibility of our AI ensemble. D.M. and V.K. optimized the HAL 
cluster---both at the hardware and software level---to maximize its throughput for 
AI training and inference at scale. B.B., I.F., and D.K. informed and guided the construction 
of the $\texttt{DLHub}\rightarrow\texttt{funcX}\rightarrow\texttt{HAL}$ workflow. 
M.L. and R.C. ran an independent AI analysis using the aforementioned workflow 
to establish the reproducibility and scalability of the results presented in this article. 
All authors contributed to developing the ideas, and writing and reviewing this manuscript.

\section*{Competing Interests}
The authors declare no competing interests.

\section*{Additional information}

\noindent \textbf{Correspondence and requests for materials} should be addressed to E.A.H.
\vspace{2mm}

\noindent \textbf{Peer review information} \textit{Nature Astronomy} thanks Elena Cuoco, Plamen 
Krastev and Linqing Wen for their contribution to the peer review of this work.
\vspace{2mm}

\noindent \textbf{Reprints and permissions information} is available at \href{http://www.nature.com/reprints}{reprints}
\vspace{2mm}

\noindent \textbf{Behind the Paper Blog} in \textit{Nature Astronomy} of this work is available at \href{https://astronomycommunity.nature.com/posts/from-disruption-to-sustained-innovation-artificial-intelligence-for-gravitational-wave-astrophysics}{From Disruption to Sustained Innovation: Artificial Intelligence for Gravitational Wave Astrophysics}

\end{document}